\documentclass[12pt,onecolumn,letterpaper]{IEEEtran}
\usepackage{epsfig}
\usepackage{amsmath}
\usepackage{amssymb}
\usepackage{algorithm}
\usepackage{algorithmic}
\usepackage{multirow}
\usepackage{url}
\usepackage[numbers]{natbib}
\usepackage[tight,footnotesize]{subfigure}
\usepackage{amsthm}
\usepackage{mathrsfs}
\usepackage[pagebackref=false,breaklinks=true,letterpaper=true,colorlinks,citecolor=blue,bookmarks=true]{hyperref}
\usepackage{enumerate}

\newtheorem{theorem}{Theorem}
\newtheorem{definition}{Definition}
\newtheorem{lemma}{Lemma}
\newtheorem{property}{Property}

\newtheorem{corollary}{Corollary}
\newtheorem{remark}{Remark}
\begin{document}

\title{Unified Scalable Equivalent Formulations for Schatten Quasi-Norms}

\author{Fanhua~Shang,~\IEEEmembership{Member,~IEEE,}
        Yuanyuan~Liu,
        and James~Cheng
\IEEEcompsocitemizethanks{\IEEEcompsocthanksitem F. Shang, Y. Liu and J. Cheng are with the Department of Computer Science and Engineering, The Chinese University of Hong Kong,
Shatin, N.T., Hong Kong. E-mail: \{fhshang, yyliu, jcheng\}@cse.cuhk.edu.hk.\protect}
\thanks{CUHK Technical Report CSE-ShangLC20160307, March 7, 2016.}}

% The paper headers
\markboth{}%
{Shell \MakeLowercase{\textit{et al.}}: Bare Demo of IEEEtran.cls for Computer Society Journals}
\maketitle

%\vspace{-10mm}
%\IEEEtitleabstractindextext{
\begin{abstract}
The Schatten quasi-norm can be used to bridge the gap between the nuclear norm and rank function, and is the tighter approximation to matrix rank. However, most existing Schatten quasi-norm minimization (SQNM) algorithms, as well as for nuclear norm minimization (NNM), are too slow or even impractical for large-scale problems, due to the singular value decomposition (SVD) or eigenvalue decomposition (EVD) of the whole matrix in each iteration. In this paper, we rigorously prove that for any $p$, $p_{1}$, $p_{2}\!>\!0$ satisfying $1/p\!=\!1/p_{1}\!+\!1/p_{2}$, the Schatten-$p$ quasi-norm of any matrix is equivalent to minimizing the product of the Schatten-$p_{1}$ norm (or quasi-norm) and Schatten-$p_{2}$ norm (or quasi-norm) of its two factor matrices. Then we present and prove the equivalence relationship between the product formula of the Schatten quasi-norm and its weighted sum formula for the two cases of $p_{1}$ and $p_{2}$: $p_{1}\!=\!p_{2}$ and $p_{1}\!\neq\!p_{2}$. In particular, when $p\!>\!1/2$, there is an equivalence between the Schatten-$p$ quasi-norm of any matrix and the Schatten-$2p$ norms of its two factor matrices, where the widely used equivalent formulation of the nuclear norm, i.e., $\|X\|_{*}\!=\!\min_{X=UV^{T}}(\|U\|^{2}_{F}\!+\!\|V\|^{2}_{F})/{2}$, can be viewed as a special case. That is, various SQNM problems with $p\!>\!1/2$ can be transformed into the one only involving smooth, convex norms of two factor matrices, which can lead to simpler and more efficient algorithms than conventional methods.

We further extend the theoretical results of two factor matrices to the cases of three and more factor matrices, from which we can see that for any $0\!<\!p\!<\!1$, the Schatten-$p$ quasi-norm of any matrix is the minimization of the mean of the Schatten-$(\lfloor1/p\rfloor\!+\!1)p$ norms of all factor matrices, where $\lfloor1/p\rfloor$ denotes the largest integer not exceeding $1/p$. In other words, for any $0\!<\!p\!<\!1$, the SQNM problem can be transformed into an optimization problem only involving the smooth, convex norms of multiple factor matrices. In addition, we also present some representative examples for two and three factor matrices. Naturally, the bi-nuclear and Frobenius/nuclear quasi-norms defined in our previous paper~\cite{shang:snm} and the tri-nuclear quasi-norm defined in our previous paper~\cite{shang:tsnm} are three important special cases.
\end{abstract}

%}

\IEEEpeerreviewmaketitle

\newpage

\section{Introduction}
\IEEEPARstart{T}{he} affine rank minimization problem arises directly in various areas of science and engineering including statistics, machine learning, information theory, data mining, medical imaging and computer vision. Some representative applications include low-rank matrix completion (LRMC)~\cite{candes:emc}, robust principal component analysis (RPCA)~\cite{candes:rpca}, low-rank representation~\cite{liu:lrr}, multivariate regression~\cite{yuan:mr}, multi-task learning~\cite{argyriou:mtl} and system identification~\cite{liu:nnm}. To efficiently solve such problems, we mainly relax the rank function to its tractable convex envelope, i.e., the nuclear norm (sum of the singular values, also known as the trace norm or Schatten-$1$ norm), which leads to a convex optimization problem~\cite{candes:emc,fazel:rmh,candes:crmc,recht:nnm}. In fact, the nuclear norm of one matrix is the $\ell_{1}$-norm of the vector of its singular values, and thus it can motivate a low-rank solution. However, it has been shown in~\citep{fan:vs,zhang:mcp} that the $\ell_{1}$-norm over-penalizes large entries of vectors, and therefore results in a solution from a possibly biased solution space. Recall from the relationship between the $\ell_{1}$-norm and nuclear norm, the nuclear norm penalty shrinks all singular values equally, which also leads to over-penalize large singular values. That is, the nuclear norm may make the solution deviate from the original solution as the $\ell_{1}$-norm does. Compared with the nuclear norm, the Schatten-${p}$ quasi-norm with $0\!<\!p\!<\!1$ is non-convex, but it can give a closer approximation to the rank function. Thus, the Schatten-${p}$ quasi-norm minimization (SQNM) has received a significant amount of attention from researchers in various communities, such as images recovery~\cite{lu:lrm}, collaborative filtering~\cite{nie:rmc,aravkin:rpca} and MRI analysis~\cite{majumdar:mri}.

Recently, two classes of iterative reweighted lease squares (IRLS) algorithms in \citep{mohan:mrm} and \citep{lai:irls} were proposed to approximate associated Schatten-${p}$ quasi-norm minimization problems, respectively. In addition, Lu et al.\ \cite{lu:lrm} proposed a family of iteratively reweighted nuclear norm (IRNN) algorithms to solve various non-convex surrogate (including the Schatten quasi-norm) minimization problems. In~\cite{lu:lrm,nie:rmc,nie:lrmr,marjanovic:mc}, the Schatten-${p}$ quasi-norm has been shown to be empirically superior to the nuclear norm for many different problems. Moreover, \cite{zhang:ncmr} theoretically proved that the SQNM requires significantly fewer measurements than conventional nuclear norm minimization (NNM). However, existing algorithms mentioned above have to be solved iteratively and involve singular value decomposition (SVD) or eigenvalue decomposition (EVD) in each iteration, as well as those for NNM. Thus they suffer from high computational cost and are even not applicable for large-scale problems~\cite{shang:snm,shang:tsnm}.

On the contrary, the nuclear norm has a scalable equivalent formulation, also known as the bilinear spectral penalty~\cite{recht:nnm,srebro:mmmf,mazumder:sr}, which has been successfully applied in many large-scale applications, such as collaborative filtering~\cite{aravkin:rpca,mitra:lsmf,recht:sgd}. In addition, Zuo et al.\ \cite{zuo:gist} proposed a generalized shrinkage-thresholding operator to iteratively solve $\ell_{p}$ quasi-norm minimization with arbitrary $p$ values, i.e., $0\!\leq\! p\!<\!1$. Since the Schatten-${p}$ quasi-norm of one matrix is equivalent to the $\ell_{p}$ quasi-norm on its singular values, we may naturally ask the following question: can we design a unified scalable equivalent formulation to the Schatten-${p}$ quasi-norm with arbitrary $p$ values, i.e., $0\!<\! p\!<\!1$.

In this paper, we first present and prove the equivalence relationship between the Schatten-$p$ quasi-norm of any matrix and the minimization of the product of the Schatten-$p_{1}$ norm (or quasi-norm) and Schatten-$p_{2}$ norm (or quasi-norm) of its two factor matrices, for any $p$, $p_{1}$, $p_{2}\!>\!0$ satisfying $1/p\!=\!1/p_{1}\!+\!1/p_{2}$. In addition, we also prove the equivalence relationship between the product formula of the Schatten quasi-norm and its weighted sum formula for the two cases of $p_{1}$ and $p_{2}$: $p_{1}\!=\!p_{2}$ and $p_{1}\!\neq\!p_{2}$. When $p\!>\!1/2$ and by setting the same value for $p_{1}$ and $p_{2}$, there is an equivalence between the Schatten-$p$ quasi-norm (or norm) of any matrix and the Schatten-$2p$ norms of its two factor matrices, where a representative example is the widely used equivalent formulation of the nuclear norm, i.e., $\|X\|_{*}\!=\!\min_{X=UV^{T}}(\|U\|^{2}_{F}\!+\!\|V\|^{2}_{F})/{2}$. In other worlds, various SQNM problems with $p\!>\!1/2$ can be transformed into the one only involving the smooth convex norms of two factor matrices, which can lead to simpler and more efficient algorithms than conventional methods~\cite{lu:lrm,nie:rmc,mohan:mrm,lai:irls,nie:lrmr,marjanovic:mc}.

We further extend the theoretical results of two factor matrices to the cases of three and more factor matrices, from which we can know that for any $0\!<\!p\!<\!1$, the Schatten-$p$ quasi-norm of any matrix is equivalent to the minimization of the mean of the Schatten-$(\lfloor1/p\rfloor\!+\!1)p$ norms of all factor matrices, where $\lfloor1/p\rfloor$ denotes the largest integer not exceeding $1/p$. Note that the norms of all factor matrices are convex and smooth. Besides the theoretical results, we also present several representative examples for two and three factor matrices. Naturally, the bi-nuclear and Frobenius/nuclear quasi-norms defined in our previous paper~\cite{shang:snm} and the tri-nuclear quasi-norm defined in our previous paper~\cite{shang:tsnm} are three important special cases.

\section{Notations and Background}
\begin{definition}\label{def1}
The Schatten-${p}$ norm ($0\!<\!p\!<\!\infty$) of a matrix $X\!\in\!\mathbb{R}^{m\times n}$ (without loss of generality, we can assume that $m\!\geq\!n$) is defined as
\begin{equation}\label{equ1}
\|X\|_{S_{p}} = \left(\sum_{i=1}^{n} \sigma^{p}_{i}(X)\right)^{1/p},
\end{equation}
where $\sigma_{i}(X)$ denotes the $i$-th singular value of $X$.
\end{definition}
When $p\!\geq\!1$, Definition~\ref{def1} defines a natural norm, for instance, the Schatten-$1$ norm is the so-called nuclear norm, $\|X\|_{*}$, and the Schatten-$2$ norm is the well-known Frobenius norm,  whereas it defines a quasi-norm for $0\!<\!p\!<\!1$. As the non-convex surrogate for the rank function, the Schatten-${p}$ quasi-norm is the better approximation than the nuclear norm~\cite{zhang:ncmr}, analogous to the superiority of the $\ell_{p}$ quasi-norm to the $\ell_{1}$-norm~\cite{lai:irls, foucart:lp}.

To recover a low-rank matrix from a small set of linear observations, $b\!\in\!\mathbb{R}^{l}$, the general SQNM problem is formulated as follows:
\begin{equation}\label{equ2}
\min_{X\in\mathbb{R}^{m\times n}}\|X\|^{p}_{S_{p}},\;\;\textrm{subject to}\;\;\mathcal{A}(X)=b
\end{equation}
where $\mathcal{A}:\mathbb{R}^{m\times n}\!\rightarrow\! \mathbb{R}^{l}$ is a general linear operator. Alternatively, the Lagrangian version of \eqref{equ2} is
\begin{equation}\label{equ3}
\min_{X\in\mathbb{R}^{m\times n}} \lambda\|X\|^{p}_{S_{p}}+f\!\left(\mathcal{A}(X)-b \right)
\end{equation}
where $\lambda\!>\!0$ is a regularization parameter, and the loss function $f(\cdot):\mathbb{R}^{l}\!\rightarrow\! \mathbb{R}$ generally denotes certain measurement for characterizing the loss $\mathcal{A}(X)-b$. For instance, $\mathcal{A}$ is the linear projection operator $\mathcal{P}_{\Omega}$, and $f(\cdot)=\|\!\cdot\!\|^{2}_{2}$ in LRMC problems~\cite{lu:lrm, mohan:mrm, marjanovic:mc, liu:nnr}, where $\mathcal{P}_{\Omega}$ is the orthogonal projection onto the linear subspace of matrices supported on $\Omega\!:=\!\{(i,j)|D_{ij}\,\textup{is observed}\}$: $\mathcal{P}_{\Omega}(D)_{ij}\!=\!D_{ij}$ if $(i,j)\!\in\!\Omega$ and $\mathcal{P}_{\Omega}(D)_{ij}\!=\!0$ otherwise. In addition, for RPCA problems~\citep{candes:rpca,shang:rpca,xu:rpca,chen:lrmr,shang:rbf}, $\mathcal{A}$ is the identity operator and $f(\cdot)\!=\!\|\!\cdot\!\|_{1}$. In the problem of multivariate regression~\citep{hsieh:nnm}, $\mathcal{A}(X)\!=\!AX$ with $A$ being a given matrix, and $f(\cdot)\!=\!\|\!\cdot\!\|^{2}_{F}$. $f(\cdot)$ may be chosen as the Hinge loss in~\citep{srebro:mmmf} or the $\ell_{p}$ quasi-norm in~\citep{nie:rmc}.

Generally, the SQNM problem, such as \eqref{equ2} and \eqref{equ3}, is non-convex, non-smooth and even non-Lipschitz~\cite{bian:ipa}. So far, only few algorithms, such as IRLS~\cite{mohan:mrm, lai:irls} and IRNN~\cite{lu:lrm}, have been developed to solve such challenging problems. However, since most existing SQNM algorithms involve SVD or EVD of the whole matrix in each iteration, they suffer from a high computational cost of $O(n^{2}m)$, which severely limits their applicability to large-scale problems~\cite{shang:snm,shang:tsnm}. While there have been many efforts towards fast SVD or EVD computation such as partial SVD~\cite{larsen:svd}, the performance of those methods is still unsatisfactory for many real applications~\cite{cai:fsvt}. As an alternative to reduce the computational complexity of SVD or EVD on a large matrix, one can factorize $X$ into two smaller factor matrices, i.e., $X\!=\!UV^{T}$. According to the unitary invariant property of norms, \eqref{equ2} and \eqref{equ3} can be reformulated into two much smaller matrices optimization problems as in~\cite{liu:as,shang:hotd}, which are still non-convex, non-smooth and even non-Lipschitz. Therefore, it is a very important problem that how to transform the challenging problems such as \eqref{equ2} and \eqref{equ3} into more tractable ones, which can be solved by simpler and more efficient algorithms.

\section{Main Results}
In this section, we first present and prove the equivalence relationship between the Schatten-$p$ quasi-norm of any matrix and the Schatten-$p_{1}$ and Schatten-$p_{2}$ quasi-norms (or norms) of its two factor matrices, where $1/p\!=\!1/p_{1}\!+\!1/p_{2}$ with any $p_{1}\!>\!0$ and $p_{2}\!>\!0$. Moreover, we prove the equivalence relationship between the product formula of the Schatten quasi-norm and its weighted sum formula for the two cases of $p_{1}$ and $p_{2}$: $p_{1}\!=\!p_{2}$ and $p_{1}\!\neq\!p_{2}$, respectively. For any $1/2\!<\!p\!\leq\!1$, the Schatten-$p$ quasi-norm (or norm) of any matrix is equivalent to the minimization of the mean of the Schatten-$2p$ norms of both factor matrices, for instance $\|X\|_{*}\!=\!\min_{X=UV^{T}}(\|U\|^{2}_{F}\!+\!\|V\|^{2}_{F})/{2}$, which can lead to simpler and more efficient algorithms than conventional methods. Finally, we extend the theoretical results of two factor matrices to the cases of three and more factor matrices. We can see that for any $0\!<\!p\!<\!1$, the Schatten-$p$ quasi-norm of any matrix is the minimization of the mean of the Schatten-$(\lfloor1/p\rfloor\!+\!1)p$ norms of all factor matrices, where $\lfloor1/p\rfloor$ denotes the largest integer not exceeding $1/p$.

\subsection{Unified Schatten Quasi-Norm Formulations of Two Factor Matrices}
\begin{theorem}\label{the11}
For any matrix $X\!\in\!\mathbb{R}^{m\times n}$ with $\textrm{rank}(X)\!=\!r\!\leq\! d$, it can be decomposed into the product of two much smaller matrices $U\!\in\!\mathbb{R}^{m\times d}$ and $V\!\in\! \mathbb{R}^{n\times d}$, i.e., $X\!=\!UV^{T}$. For any $0\!<\!p\!\leq\!1$, $p_{1}\!>\!0$ and $p_{2}\!>\!0$ satisfying $1/p_{1}\!+\!1/p_{2}\!=\!1/p$, then
\begin{equation}\label{eqthe1}
\|X\|_{S_{p}}=\min_{U\in\mathbb{R}^{m\times d},V\in\mathbb{R}^{n\times d}:X=UV^{T}}\|U\|_{S_{p_{1}}}\|V\|_{S_{p_{2}}}.
\end{equation}
\end{theorem}

The detailed proof of Theorem~\ref{the11} is provided in Section~\ref{proofs1}. From Theorem~\ref{the11}, it is very clear that for any $0\!<\!p\!\leq\!1$ and $p_{1},p_{2}\!>\!0$ satisfying ${1}/{p}\!=\!{1}/{p_{1}}\!+\!{1}/{p_{2}}$, then the Schatten-$p$ quasi-norm (or norm) of any matrix $X$ is equivalent to minimizing the product of the Schatten-$p_{1}$ norm (or quasi-norm) and Schatten-$p_{2}$ norm (or quasi-norm) of its two factor matrices.

Naturally, we can see that $p_{1}$ and $p_{2}$ may have the same value, i.e., $p_{1}\!=\!p_{2}\!=\!2p$, or different values, i.e., $p_{1}\!\neq\! p_{2}$. Next, we discuss these two cases for $p_{1}$ and $p_{2}$, i.e., $p_{1}\!=\!p_{2}$ and $p_{1}\!\neq\! p_{2}$.

\subsubsection{Case of $p_{1}\!=\!p_{2}$}
First, we discuss the case when $p_{1}\!=\!p_{2}$. In fact, for any given $0\!<\!p\!\leq\!1$, there exist infinitely many pairs of positive numbers $p_{1}$ and $p_{2}$ satisfying ${1}/{p_{1}}\!+\!{1}/{p_{2}}\!=\!{1}/{p}$, such that the equality~\eqref{eqthe1} holds. By setting the same value for $p_{1}$ and $p_{2}$, i.e., $p_{1}\!=\!p_{2}\!=\!2p$, we give a unified scalable equivalent formulation for the Schatten-$p$ quasi-norm (or norm) as follows.

\begin{theorem}\label{cor1}
Given any matrix $X\!\in\!\mathbb{R}^{m\times n}$ of $\textrm{rank}(X)\!=\!r\!\leq\! d$, then the following equalities hold:
\begin{equation}\label{eqthe2}
\boxed{%
\begin{split}
\|X\|_{S_{p}}&=\min_{U\in\mathbb{R}^{m\times d},V\in\mathbb{R}^{n\times d}:X=UV^{T}}\|U\|_{S_{2p}}\|V\|_{S_{2p}}\\
&=\min_{U\in\mathbb{R}^{m\times d},V\in\mathbb{R}^{n\times d}:X=UV^{T}}\left(\frac{\|U\|^{2p}_{S_{2p}}\!+\|V\|^{2p}_{S_{2p}}}{2}\right)^{1/p}.
\end{split}
}
\end{equation}
\end{theorem}

\begin{remark}
The detailed proof of Theorem~\ref{cor1} is provided in Section~\ref{proofs2}. From the second equality in \eqref{eqthe2}, we know that, for any $0\!<\!p\!\leq\!1$, the Schatten-$p$ quasi-norm (or norm) minimization problems in many low-rank matrix completion and recovery applications can be transformed into the one of minimizing the mean of the Schatten-$2p$ norms (or quasi-norms) of both much smaller factor matrices. We note that when $1/2\!<\!p\!\leq\!1$, the norms of both much smaller factor matrices are convex and smooth (see Example 2 below) due to $2p\!>\!1$, which can lead to simpler and more efficient algorithms than conventional methods~\cite{lu:lrm,nie:rmc,mohan:mrm,lai:irls,nie:lrmr,marjanovic:mc}.
\end{remark}

When $p\!=\!1$ and $p_{1}\!=\!p_{2}\!=\!2$, the equalities in Theorem~\ref{cor1} become the following forms.
\begin{corollary}\label{cor2}
Given any matrix $X\!\in\!\mathbb{R}^{m\times n}$ with $\textrm{rank}(X)\!=\!r\!\leq\! d$, the following equalities hold:
\begin{equation}\label{eqthe3}
\begin{split}
\|X\|_{*}&=\min_{U\in\mathbb{R}^{m\times d},V\in\mathbb{R}^{n\times d}:X=UV^{T}}\|U\|_{F}\|V\|_{F}\\
&=\min_{U\in\mathbb{R}^{m\times d},V\in\mathbb{R}^{n\times d}:X=UV^{T}}\frac{\|U\|^{2}_{F}+\|V\|^{2}_{F}}{2}.
\end{split}
\end{equation}
\end{corollary}

The bilinear spectral penalty in the second equality of~\eqref{eqthe3} has been widely used in many low-rank matrix completion and recovery problems, such as collaborative filtering~\cite{recht:nnm, srebro:mmmf}, RPCA~\cite{cabral:nnbf}, online RPCA~\cite{feng:orpca}, and image recovery~\cite{jin:hankel}. Note that the well-known equivalent formulations of the nuclear norm in Corollary~\ref{cor2} are just a special case of Theorem~\ref{cor1}, i.e., $p\!=\!1$ and $p_{1}\!=\!p_{2}\!=\!2$. In the following, we give two more representative examples for the case of $p_{1}\!=\!p_{2}$.

\textbf{Example 1}: When $p\!=\!1/2$, and by setting $p_{1}\!=\!p_{2}\!=\!1$ and using Theorem~\ref{the11}, we have
\begin{equation*}
\|X\|_{S_{1/2}}=\min_{U\in\mathbb{R}^{m\times d},V\in\mathbb{R}^{n\times d}:X=UV^{T}}\|U\|_{*}\|V\|_{*}.
\end{equation*}
Due to the basic inequality $xy\leq (\frac{x+y}{2})^2$ for any real numbers $x$ and $y$, we obtain
\begin{equation*}
\begin{split}
\|X\|_{S_{1/2}}&=\min_{U\in\mathbb{R}^{m\times d},V\in\mathbb{R}^{n\times d}:X=UV^{T}}\|U\|_{*}\|V\|_{*}\\
&\leq\min_{U\in\mathbb{R}^{m\times d},V\in\mathbb{R}^{n\times d}:X=UV^{T}}\left(\frac{\|U\|_{*}+\|V\|_{*}}{2}\right)^{2}.
\end{split}
\end{equation*}
Let $U_{\star}\!=\!L_{X}\Sigma^{1/2}_{X}$ and $V_{\star}\!=\!R_{X}\Sigma^{1/2}_{X}$ as in~\cite{shang:snm,shang:tsnm}, then we have $X\!=\!U_{\star}V^{T}_{\star}$ and
\begin{equation*}
\|X\|_{S_{1/2}}=\left(\textup{Tr}^{1/2}(\Sigma_{X})\right)^{2}=\|U_{\star}\|_{*}\|V_{\star}\|_{*}=\left(\frac{\|U_{\star}\|_{*}+\|V_{\star}\|_{*}}{2}\right)^{2}
\end{equation*}
where $\textup{Tr}^{1/2}(\Sigma_{X})\!=\!\sum_{i}(\Sigma_{X})^{1/2}_{i,i}$. Therefore, under the constraint $X\!=\!UV^{T}$, we have the following property~\cite{shang:snm,shang:tsnm}.
\begin{property}\label{pro1}
\begin{equation}\label{eqthe4}
\begin{split}
\|X\|_{S_{1/2}}&=\min_{U\in\mathbb{R}^{m\times d},V\in\mathbb{R}^{n\times d}:X=UV^{T}}\|U\|_{*}\|V\|_{*}\\
&=\min_{U\in\mathbb{R}^{m\times d},V\in\mathbb{R}^{n\times d}:X=UV^{T}}\left(\frac{\|U\|_{*}+\|V\|_{*}}{2}\right)^{2}.
\end{split}
\end{equation}
\end{property}

In our previous papers~\cite{shang:snm,shang:tsnm}, the scalable formulations in the above equalities are known as the bi-nuclear quasi-norm. In other words, the bi-nuclear quasi-norm is also a special case of Theorem~\ref{cor1}, i.e., $p\!=\!1/2$ and $p_{1}\!=\!p_{2}\!=\!1$.

\textbf{Example 2}: When $p\!=\!2/3$, and by setting $p_{1}\!=\!p_{2}\!=\!4/3$ and using Theorem~\ref{the11}, we have
\begin{equation*}
\|X\|_{S_{2/3}}=\min_{U\in\mathbb{R}^{m\times d},V\in\mathbb{R}^{n\times d}:X=UV^{T}}\|U\|_{S_{4/3}}\|V\|_{S_{4/3}}.
\end{equation*}
Due to the basic inequality $xy\leq \frac{x^2+y^2}{2}$ for any real numbers $x$ and $y$, then
\begin{equation*}
\begin{split}
\|X\|_{S_{2/3}}&=\min_{U\in\mathbb{R}^{m\times d},V\in\mathbb{R}^{n\times d}:X=UV^{T}}\|U\|_{S_{4/3}}\|V\|_{S_{4/3}}\\
&=\min_{U\in\mathbb{R}^{m\times d},V\in\mathbb{R}^{n\times d}:X=UV^{T}}\left(\|U\|^{2/3}_{S_{4/3}}\|V\|^{2/3}_{S_{4/3}}\right)^{3/2}\\
&\leq\min_{U\in\mathbb{R}^{m\times d},V\in\mathbb{R}^{n\times d}:X=UV^{T}}\left(\frac{\|U\|^{4/3}_{S_{4/3}}\!+\|V\|^{4/3}_{S_{4/3}}}{2}\right)^{3/2}.
\end{split}
\end{equation*}

Let $U_{\star}\!=\!L_{X}\Sigma^{1/2}_{X}$ and $V_{\star}\!=\!R_{X}\Sigma^{1/2}_{X}$, then we have $X\!=\!U_{\star}V^{T}_{\star}$ and
\begin{equation*}
\|X\|_{S_{2/3}}=\left(\textup{Tr}^{2/3}(\Sigma_{X})\right)^{3/2}=\|U_{\star}\|_{S_{4/3}}\|V_{\star}\|_{S_{4/3}}=\left(\frac{\|U_{\star}\|^{4/3}_{S_{4/3}}\!+\|V_{\star}\|^{4/3}_{S_{4/3}}}{2}\right)^{3/2}.
\end{equation*}
Together with the constraint $X\!=\!UV^{T}$, thus we have the following property.
\begin{property}\label{pro2}
\begin{equation}\label{eqthe5}
\begin{split}
\|X\|_{S_{2/3}}&=\min_{U\in\mathbb{R}^{m\times d},V\in\mathbb{R}^{n\times d}:X=UV^{T}}\|U\|_{S_{4/3}}\|V\|_{S_{4/3}}\\
&=\min_{U\in\mathbb{R}^{m\times d},V\in\mathbb{R}^{n\times d}:X=UV^{T}}\left(\frac{\|U\|^{4/3}_{S_{4/3}}\!+\|V\|^{4/3}_{S_{4/3}}}{2}\right)^{3/2}.
\end{split}
\end{equation}
\end{property}

\subsubsection{Case of $p_{1}\!\neq\! p_{2}$}
In this part, we discuss the case of $p_{1}\!\neq\! p_{2}$. Different from the case of $p_{1}\!=\! p_{2}$, we may set infinitely many different values for $p_{1}$ and $p_{2}$. For any given $0\!<\!p\!\leq\!1$, there must exist $p_{1},p_{2}\!>\!0$, at least one of which is no less than 1 (which means that the norm of one factor matrix can be convex), such that ${1}/{p}\!=\!{1}/{p_{1}}\!+\!{1}/{p_{2}}$. Indeed, for any $0\!<\!p\!\leq\!1$, the values of $p_{1}$ and $p_{2}$ may be different, e.g., $p_{1}\!=\!1$ and $p_{2}\!=\!2$ for $p\!=\!2/3$, thus we give the following unified scalable equivalent formulations for the Schatten-$p$ quasi-norm (or norm).

\begin{theorem}\label{cor3}
Given any matrix $X\!\in\!\mathbb{R}^{m\times n}$ of $\textrm{rank}(X)\!=\!r\leq\! d$, and any $0\!<\!p\!\leq\!1$, $p_{1}\!>\!0$ and $p_{2}\!>\!0$ satisfying ${1}/{p_{1}}\!+\!{1}/{p_{2}}\!=\!{1}/{p}$, then the following equalities hold:
\begin{equation}\label{eqthe61}
\boxed{%
\begin{split}
\|X\|_{S_{p}}&=\min_{U\in\mathbb{R}^{m\times d},V\in\mathbb{R}^{n\times d}:X=UV^{T}}\|U\|_{S_{p_{1}}}\|V\|_{S_{p_{2}}}\\
&=\min_{U\in\mathbb{R}^{m\times d},V\in\mathbb{R}^{n\times d}:X=UV^{T}}\left(\frac{p_{2}\|U\|^{p_{1}}_{S_{p_{1}}}\!+p_{1}\|V\|^{p_{2}}_{S_{p_{2}}}}{p_{1}+p_{2}}\right)^{1/p}\\
&=\min_{U\in\mathbb{R}^{m\times d},V\in\mathbb{R}^{n\times d}:X=UV^{T}}\left(\frac{\|U\|^{p_{1}}_{S_{p_{1}}}/p_{1}\!+\|V\|^{p_{2}}_{S_{p_{2}}}/p_{2}}{1/p}\right)^{1/p}.
\end{split}
}
\end{equation}
\end{theorem}

\begin{remark}
The detailed proof of Theorem~\ref{cor3} is given in Section~\ref{proofs3}. From Theorem~\ref{cor3}, we know that Theorem~\ref{cor1} and Corollary~\ref{cor2} can be viewed as two special cases of Theorem~\ref{cor3}, i.e., $p_{1}\!=\!p_{2}\!=\!2p$ and $p_{1}\!=\!p_{2}\!=\!2$, respectively. That is, Theorem~\ref{cor3} is the more general form of Theorem~\ref{cor1} and Corollary~\ref{cor2}. From the second equality in \eqref{eqthe61}, we can see that, for any $0\!<\!p\!\leq\!1$, the Schatten-$p$ quasi-norm (or norm) minimization problem can be transformed into the one of minimizing the weighted sum of the Schatten-$p_{1}$ norm (or quasi-norm) and Schatten-$p_{2}$ norm (or quasi-norm) of two much smaller factor matrices (see Example 3 and Example 4 below), where the weights of the two terms in the second equality of \eqref{eqthe61} are $p_{2}/(p_{1}\!+\!p_{2})$ and $p_{1}/(p_{1}\!+\!p_{2})$, respectively.
\end{remark}

In the following, we give two representative examples for the case of $p_{1}\!\neq\!p_{2}$.

\textbf{Example 3}: When $p\!=\!2/3$, and by setting $p_{1}\!=\!1$ and $p_{2}\!=\!2$, and using Theorem~\ref{the11}, then
\begin{equation*}
\|X\|_{S_{2/3}}=\min_{U\in\mathbb{R}^{m\times d},V\in\mathbb{R}^{n\times d}:X=UV^{T}}\|U\|_{*}\|V\|_{F}.
\end{equation*}
In addition, we have
\begin{equation*}
\begin{split}
\|U\|_{*}\|V\|_{F}&=\sqrt{\|U\|_{*}}\sqrt{\|U\|_{*}}\|V\|_{F}\\
&^{a}\!\!\leq \left( \frac{\sqrt{\|U\|_{*}}+\sqrt{\|U\|_{*}}+\|V\|_{F}}{3}\right)^{3}\\
&= \left( \frac{2\sqrt{\|U\|_{*}}+\sqrt{\|V\|^{2}_{F}}}{3}\right)^{3}\\
&^{b}\!\!\leq \left(\frac{2\|U\|_{*}+\|V\|^{2}_{F}}{3}\right)^{{3}/{2}}
\end{split}
\end{equation*}
where the inequality $^{a}\!\!\leq$ holds due to the fact that $x_{1}x_{2}x_{3}\!\leq\![(x_{1}\!+\!x_{2}\!+\!x_{3})/3]^{3}$ for any real numbers $x_{1}$, $x_{2}$ and $x_{3}$, and the inequality $^{b}\!\!\leq$ follows from the Jensen's inequality for the concave function $g(x)\!=\!x^{1/2}$.

Let $U_{\star}\!=\!L_{X}\Sigma^{2/3}_{X}$ and $V_{\star}\!=\!R_{X}\Sigma^{1/3}_{X}$ as in~\cite{shang:snm}, then we have $X\!=\!U_{\star}V^{T}_{\star}$ and
\begin{equation*}
\|X\|_{S_{2/3}}=\left(\textup{Tr}^{2/3}(\Sigma_{X})\right)^{3/2}=\|U_{\star}\|_{*}\|V_{\star}\|_{F}=\left(\frac{2\|U_{\star}\|_{*}+\|V_{\star}\|^{2}_{F}}{3}\right)^{{3}/{2}}.
\end{equation*}
Therefore, together with the constraint $X\!=\!UV^{T}$, we have the following property~\cite{shang:snm}.
\begin{property}\label{pro3}
\begin{equation}\label{eqthe6}
\begin{split}
\|X\|_{S_{2/3}}=&\min_{U\in\mathbb{R}^{m\times d},V\in\mathbb{R}^{n\times d}:X=UV^{T}}\|U\|_{*}\|V\|_{F}\\
=&\min_{U\in\mathbb{R}^{m\times d},V\in\mathbb{R}^{n\times d}:X=UV^{T}}\!\left(\frac{2\|U\|_{*}+\|V\|^{2}_{F}}{3}\right)^{3/2}.
\end{split}
\end{equation}
\end{property}
In our previous paper~\cite{shang:snm}, the scalable formulations in the above equalities are known as the Frobenius/nuclear hybrid quasi-norm. It is clear that the Frobenius/nuclear hybrid quasi-norm is also a special case of Theorem~\ref{cor3}, i.e., $p\!=\!2/3$, $p_{1}\!=\!1$ and $p_{2}\!=\!2$. As shown in the above representative examples and our previous papers~\cite{shang:snm,shang:tsnm}, we can design more efficient algorithms to solve the Schatten-$p$ quasi-norm with $1/2\!\leq\! p\!<\!1$ than conventional methods~\cite{lu:lrm,nie:rmc,mohan:mrm,lai:irls,nie:lrmr,marjanovic:mc}.

\textbf{Example 4}: When $p\!=\!2/5$, and by setting $p_{1}\!=\!1/2$ and $p_{2}\!=\!2$, and using Theorem~\ref{the11}, we have
\begin{equation*}
\|X\|_{S_{2/5}}=\min_{U\in\mathbb{R}^{m\times d},V\in\mathbb{R}^{n\times d}:X=UV^{T}}\|U\|_{S_{1/2}}\|V\|_{F}.
\end{equation*}
Moreover,
\begin{equation*}
\begin{split}
\|U\|_{S_{1/2}}\|V\|_{F}&=\left(\|U\|^{1/4}_{S_{1/2}}\right)^{4}\|V\|_{F}\\
&^{a}\!\!\leq \left(\frac{4\sqrt{\|U\|^{1/2}_{S_{1/2}}}\!+\sqrt{\|V\|^{2}_{F}}}{5}\right)^{5}\\
&^{b}\!\!\leq \left(\frac{4\|U\|^{1/2}_{S_{1/2}}\!+\|V\|^{2}_{F}}{5}\right)^{5/{2}}
\end{split}
\end{equation*}
where the inequality $^{a}\!\!\leq$ holds due to the familiar inequality of arithmetic and geometric means, and the inequality $^{b}\!\!\leq$ follows from the Jensen's inequality for the concave function $g(x)\!=\!x^{1/2}$.

Let $U_{\star}\!=\!L_{X}\Sigma^{4/5}_{X}$ and $V_{\star}\!=\!R_{X}\Sigma^{1/5}_{X}$, then we have $X\!=\!U_{\star}V^{T}_{\star}$ and
\begin{equation*}
\|X\|_{S_{2/5}}=\left(\textup{Tr}^{2/5}(\Sigma_{X})\right)^{5/2}=\|U_{\star}\|_{S_{1/2}}\|V_{\star}\|_{F}=\left(\frac{4\|U_{\star}\|^{1/2}_{S_{1/2}}\!+\|V_{\star}\|^{2}_{F}}{5}\right)^{{5}/{2}}.
\end{equation*}
With the constraint $X\!=\!UV^{T}$, thus we have the following property.
\begin{property}\label{pro4}
\begin{equation}\label{eqthe7}
\begin{split}
\|X\|_{S_{2/5}}=&\min_{U\in\mathbb{R}^{m\times d},V\in\mathbb{R}^{n\times d}:X=UV^{T}}\|U\|_{S_{1/2}}\|V\|_{F}\\
=&\min_{U\in\mathbb{R}^{m\times d},V\in\mathbb{R}^{n\times d}:X=UV^{T}}\!\left(\frac{4\|U\|^{1/2}_{S_{1/2}}\!+\|V\|^{2}_{F}}{5}\right)^{5/2}.
\end{split}
\end{equation}
\end{property}

\subsection{Extensions to Multiple Factor Matrices}

\begin{theorem}\label{the12}
For any matrix $X\!\in\!\mathbb{R}^{m\times n}$ of $\textrm{rank}(X)\!=\!r\!\leq\! d$, it can be decomposed into the product of three much smaller matrices $U\!\in\!\mathbb{R}^{m\times d}$, $V\!\in\! \mathbb{R}^{d\times d}$ and $W\!\in\! \mathbb{R}^{n\times d}$, i.e., $X\!=\!UVW^{T}$. For any $0\!<\!p\!\leq\!1$ and $p_{i}>0$ for all $i=1,2,3$, satisfying $1/p_{1}\!+\!1/p_{2}\!+\!1/p_{3}\!=\!1/p$, then
\begin{equation}\label{eqthe8}
\boxed{%
\begin{split}
\|X\|_{S_{p}}&=\min_{U\in\mathbb{R}^{m\times d},V\in\mathbb{R}^{d\times d},W\in\mathbb{R}^{n\times d}:X=UVW^{T}}\|U\|_{S_{p_{1}}}\|V\|_{S_{p_{2}}}\|W\|_{S_{p_{3}}}\\
&=\min_{U\in\mathbb{R}^{m\times d},V\in\mathbb{R}^{d\times d},W\in\mathbb{R}^{n\times d}:X=UVW^{T}} \left(\frac{p_{2}p_{3}\|U\|^{p_{1}}_{S_{p_{1}}}\!\!+p_{1}p_{3}\|V\|^{p_{2}}_{S_{p_{2}}}\!\!+p_{1}p_{2}\|W\|^{p_{3}}_{S_{p_{3}}}}{p_{2}p_{3}+p_{1}p_{3}+p_{1}p_{2}}\right)^{1/p}\\
&=\min_{U\in\mathbb{R}^{m\times d},V\in\mathbb{R}^{d\times d},W\in\mathbb{R}^{n\times d}:X=UVW^{T}} \left(\frac{\|U\|^{p_{1}}_{S_{p_{1}}}\!/p_{1}+\|V\|^{p_{2}}_{S_{p_{2}}}\!/p_{2}+\|W\|^{p_{3}}_{S_{p_{3}}}\!/p_{3}}{1/p}\right)^{1/p}.
\end{split}
}
\end{equation}
\end{theorem}

The detailed proof of Theorem~\ref{the12} is provided in Section~\ref{proofs4}. From Theorem~\ref{the12}, we can see that for any $0\!<\!p\!\leq\!1$ and $p_{1},p_{2},p_{3}\!>\!0$ satisfying ${1}/{p_{1}}\!+\!{1}/{p_{2}}\!+\!{1}/{p_{3}}\!=\!{1}/{p}$, the Schatten-$p$ quasi-norm (or norm) of any matrix is equivalent to minimizing the weighted sum of the Schatten-$p_{1}$ norm (or quasi-norm), Schatten-$p_{2}$ norm (or quasi-norm) and Schatten-$p_{3}$ norm (or quasi-norm) of these three much smaller factor matrices, where the weights of the three terms are $p/p_{1}$, $p/p_{2}$ and $p/p_{3}$, respectively. Similarly, we extend Theorem~\ref{the12} to the case of more factor matrices as follows.

\begin{theorem}\label{the13}
For any matrix $X\!\in\!\mathbb{R}^{m\times n}$ of $\textrm{rank}(X)\!=\!r\!\leq\! d$, it can be decomposed into the product of multiple much smaller matrices $U_{i}$, $i=1,2,\ldots,M$, i.e., $X\!=\!\prod^{M}_{i=1}U_{i}$. For any $0\!<\!p\!\leq\!1$ and $p_{i}>0$ for all $i=1,2,\ldots,M$, satisfying $\sum^{M}_{i=1}1/p_{i}\!=\!1/p$, then
\begin{equation}\label{eqthe81}
\boxed{%
\begin{split}
\|X\|_{S_{p}}&=\min_{U_{i}:X=\prod^{M}_{i=1}U_{i}}\prod^{M}_{i=1}\|U_{i}\|_{S_{p_{i}}}\\
&=\min_{U_{i}:X=\prod^{M}_{i=1}U_{i}}\left(\frac{\sum^{M}_{i=1}\|U_{i}\|^{p_{i}}_{S_{p_{i}}}/{p_{i}}}{1/p}\right)^{1/p}.
\end{split}
}
\end{equation}
\end{theorem}

The proof of Theorem~\ref{the13} is very similar to that of Theorem~\ref{the12} and is thus omitted. From Theorem~\ref{the13}, we can know that for any $0\!<\!p\!\leq\!1$ and $p_{i}>0$ for all $i=1,2,\ldots,M$, satisfying $\sum^{M}_{i=1}1/p_{i}\!=\!1/p$, the Schatten-$p$ quasi-norm (or norm) of any matrix is equivalent to the minimization of the weighted sum of the Schatten-$p_{i}$ norm (or quasi-norm) of each much smaller factor matrix, where the weights for these terms are $p/p_{i}$ for all $i=1,2,\ldots,M$.

Similar to the case of two factor matrices, for any given $0\!<\!p\!\leq\!1$, there exist infinitely many positive numbers $p_{1}$, $p_{2}$ and $p_{3}$ such that ${1}/{p_{1}}\!+\!{1}/{p_{2}}\!+\!{1}/{p_{3}}\!=\!{1}/{p}$, and the equality~\eqref{eqthe8} holds. By setting the same value for $p_{1}$, $p_{2}$ and $p_{3}$, i.e., $p_{1}\!=\!p_{2}\!=\!p_{3}=3p$, we give the following unified scalable equivalent formulations for the Schatten-$p$ quasi-norm (or norm).

\begin{corollary}\label{cor4}
Given any matrix $X\!\in\!\mathbb{R}^{m\times n}$ of $\textrm{rank}(X)\!=\!r\!\leq\! d$, then the following equalities hold:
\begin{equation}\label{eqthe9}
\boxed{%
\begin{split}
\|X\|_{S_{p}}&=\min_{U\in\mathbb{R}^{m\times d},V\in\mathbb{R}^{d\times d},W\in\mathbb{R}^{n\times d}:X=UVW^{T}}\|U\|_{S_{3p}}\|V\|_{S_{3p}}\|W\|_{S_{3p}}\\
&=\min_{U\in\mathbb{R}^{m\times d},V\in\mathbb{R}^{d\times d},W\in\mathbb{R}^{n\times d}:X=UVW^{T}}\left(\frac{\|U\|^{3p}_{S_{3p}}\!+\|V\|^{3p}_{S_{3p}}\!+\|W\|^{3p}_{S_{3p}}}{3}\right)^{1/p}.
\end{split}
}
\end{equation}
\end{corollary}

\begin{remark}
The detailed proof of Corollary~\ref{cor4} is provided in Section~\ref{proofs5}. From the second equality in \eqref{eqthe9}, we know that, for any $0\!<\!p\!<\!1$, various Schatten-$p$ quasi-norm minimization problems in many low-rank matrix completion and recovery applications can be transformed into the problem of minimizing the mean of the Schatten-$3p$ norms (or quasi-norms) of three much smaller factor matrices. In addition, we note that when $1/3\!<\!p\!\leq\!1$, the norms of the three factor matrices are convex and smooth due to $3p\!>\!1$, which can also lead to some simpler and more efficient algorithms than conventional methods.
\end{remark}

\textbf{Example 5}: In the following, we give a representative example. When $p\!=\!1/3$ and $p_{1}\!=\!p_{2}\!=\!p_{3}\!=\!1$, the equalities in Corollary~\ref{cor4} become the following forms~\cite{shang:tsnm}.

\begin{property}\label{pro5}
For any matrix $X\!\in\! \mathbb{R}^{m\times n}$ of $\textrm{rank}(X)\!=\!r\!\leq\! d$, then the following equalities hold:
\begin{equation}\label{eqthe10}
\begin{split}
\|X\|_{S_{1/3}}=&\min_{U\in\mathbb{R}^{m\times d},V\in\mathbb{R}^{d\times d},W\in\mathbb{R}^{n\times d}:X=UVW^{T}}\|U\|_{*}\|V\|_{*}\|W\|_{*}\\
=&\min_{U,V,W:X=UVW^{T}}\!\left(\frac{\|U\|_{*}+\|V\|_{*}+\|W\|_{*}}{3}\right)^{3}.
\end{split}
\end{equation}
\end{property}

From Property~\ref{pro5}, we can see that the tri-nuclear quasi-norm defined in our previous paper~\cite{shang:tsnm} is also a special case of Corollary~\ref{cor4}.

\textcolor[rgb]{1.00,0.00,0.00}{From Theorem~\ref{cor1}, we can know that for any $1/2\!<\!p\!\leq\!1$, the Schatten-$p$ quasi-norm (or norm) of any matrix is equivalent to minimizing the mean of the Schatten-$2p$ norms of both factor matrices, as well as Corollary~\ref{cor4} for any $1/3\!<\!p\!\leq\!1$. In other worlds, if $1/2\!<\!p\!\leq\!1$ or $1/3\!<\!p\!\leq\!1$, the original Schatten-$p$ quasi-norm (or norm) minimization problem can be transformed into a simpler one only involving the convex and smooth norms of two or three factor matrices. In addition, we extend the results of Theorem~\ref{cor1} and Corollary~\ref{cor4} to the case of more factor matrices, as shown in Corollary~\ref{cor5} below. The proof of Corollary~\ref{cor5} is very similar to that of Corollary~\ref{cor4} and is thus omitted. In other words, for any $0\!<\!p\!<\!1$, the Schatten-$p$ quasi-norm of any matrix can theoretically be equivalent to the minimization of the mean of the Schatten-$(Mp)$ norms of all $M$ factor matrices, where $M\!=\!(\lfloor1/p\rfloor\!+\!1)$ and $\lfloor1/p\rfloor$ denotes the largest integer not exceeding $1/p$. It needs to be strongly emphasized that the norms of all factor matrices are convex and smooth due to $Mp>1$, which can help us to design simpler and more efficient algorithms.
}

\begin{corollary}\label{cor5}
Given any matrix $X\!\in\!\mathbb{R}^{m\times n}$ of $\textrm{rank}(X)\!=\!r\!\leq\! d$, then the following equalities hold:
\begin{equation}\label{eqthe11}
\boxed{%
\begin{split}
\|X\|_{S_{p}}&=\min_{U_{i}:X=\prod^{M}_{i=1}U_{i}}\prod^{M}_{i=1}\|U_{i}\|_{S_{Mp}}\\
&=\min_{U_{i}:X=\prod^{M}_{i=1}U_{i}}\left(\frac{\sum^{M}_{i=1}\|U_{i}\|^{Mp}_{S_{Mp}}}{M}\right)^{1/p}.
\end{split}
}
\end{equation}
\end{corollary}

\section{Proofs}
\label{proofs}
In this section, we give the detailed proofs for some important theorems and corollaries. We first introduce several important inequalities, such as the Jensen's inequality, H\"{o}lder's inequality and Young's inequality, that we use throughout our proofs.

\begin{lemma}[Jensen's inequality]\label{lem11}
Assume that the function $g:\mathbb{R}^{+}\rightarrow \mathbb{R}^{+}$ is a continuous concave function on $[0,+\infty)$. For all $t_{i}\geq 0$ satisfying $\sum_{i}t_{i}=1$, and any $x_{i}\in \mathbb{R}^{+}$ for $i=1,\ldots,n$, then
\begin{equation}\label{dtreq1}
g\left(\sum^{n}_{i=1}t_{i}x_{i}\right)\geq \sum^{n}_{i=1}t_{i}g(x_{i}).
\end{equation}
\end{lemma}

\begin{lemma}[H\"{o}lder's inequality]\label{lem2}
For any $p,q>1$ satisfying $1/p+1/q=1$, then for any $x_{i}$ and $y_{i}$, $i=1,\ldots,n$,
\begin{equation}\label{dtreq12}
\sum^{n}_{i=1}|x_{i}y_{i}|\leq\left(\sum^{n}_{i=1}|x_{i}|^{p}\right)^{1/p}\left(\sum^{n}_{i=1}|y_{i}|^{q}\right)^{1/q}
\end{equation}
with equality iff there is a constant $c\neq0$ such that each $x^{p}_{i}=cy^{q}_{i}$.
\end{lemma}

\begin{lemma}[Young's inequality]\label{lem3}
Let $a,b\!\geq\!0$ and $1\!<\!p,q\!<\!\infty$ be such that $1/p+1/q=1$. Then
\begin{equation}\label{dtreq13}
\frac{a^{p}}{p}+\frac{b^{q}}{q}\geq ab
\end{equation}
with equality iff $a^{p}=b^{q}$.
\end{lemma}

\subsection{Proof of Theorem~\ref{the11}}
\label{proofs1}
Before giving a complete proof for Theorem~\ref{the11}, we first present and prove the following lemma.

\begin{lemma}\label{lem12}
Suppose that $Z\!\in\!\mathbb{R}^{m\times n}$ is a matrix of rank $r\leq\min(m,\,n)$, and we denote its thin SVD by $Z\!=\!L_{Z}\Sigma_{Z}R^{T}_{Z}$, where $L_{Z}\!\in\! \mathbb{R}^{m\times r}$, $R_{Z}\in \mathbb{R}^{n\times r}$ and $\Sigma_{Z}\!\in\! \mathbb{R}^{r\times r}$. For any $A\!\in\!\mathbb{R}^{r\times r}$ satisfying $AA^{T}=A^{T}A=I_{r\times r}$, and the given $p\;(0<p\leq1)$, then $(A\Sigma_{Z}A^{T})_{k,k}\geq 0$ for all $k=1,\ldots,r$, and
\begin{equation*}
\textup{Tr}^{p}(A\Sigma_{Z}A^{T})\geq \textup{Tr}^{p}(\Sigma_{Z})=\|Z\|^{p}_{S_{p}},
\end{equation*}
where $\textup{Tr}^{p}(B)=\sum_{i}B^{p}_{ii}$.
\end{lemma}

\begin{IEEEproof}
For any $k\in\{1,\ldots,r\}$, we have $(A\Sigma_{Z}A^{T})_{k,k}=\sum_{i}a^{2}_{ki}\sigma_{i}\geq 0$, where $\sigma_{i}\geq 0$ is the $i$-th singular value of $Z$. Then
\begin{equation}\label{Seq1}
\textup{Tr}^{p}(A\Sigma_{Z}A^{T})=\sum_{k}\left(\sum_{i}a^{2}_{ki}\sigma_{i}\right)^{p}.
\end{equation}

Recall that $g(x)=x^{p}$ with $0<p<1$ is a concave function on $\mathbb{R}^{+}$. By using the Jensen's inequality~\citep{mitrinovic:ai}, as stated in Lemma~\ref{lem11}, and $\sum_{i}a^{2}_{ki}=1$ for any $k\in\{1,\ldots,r\}$, we have
\begin{equation*}
\left(\sum_{i}a^{2}_{ki}\sigma_{i}\right)^{p}\geq \sum_{i}a^{2}_{ki}\sigma^{p}_{i}.
\end{equation*}

Using the above inequality and $\sum_{k}a^{2}_{ki}=1$ for any $i\in\{1,\ldots,r\}$, \eqref{Seq1} can be rewritten as
\begin{equation}\label{Seq2}
\begin{split}
\textup{Tr}^{p}(A\Sigma_{Z}A^{T}) &= \sum_{k}\left(\sum_{i}a^{2}_{ki}\sigma_{i}\right)^{p} \\
&\geq \sum_{k}\sum_{i}a^{2}_{ki}\sigma^{p}_{i}\\
&=\sum_{i}\sigma^{p}_{i}\\
&=\textup{Tr}^{p}(\Sigma_{Z})=\|Z\|^{p}_{S_{p}}.
\end{split}
\end{equation}

In addition, when $g(x)=x$, i.e., $p=1$, we obtain
\begin{equation*}
\left(\sum_{i}a^{2}_{ki}\sigma_{i}\right)^{p}= \sum_{i}a^{2}_{ki}\sigma_{i},
\end{equation*}
which means that the inequality \eqref{Seq2} is still satisfied. This completes the proof.
\end{IEEEproof}

\textbf{Proof of Theorem~\ref{the11}:}

\begin{IEEEproof}
Let $U\!=\!L_{U}\Sigma_{U}R^{T}_{U}$ and $V\!=\!L_{V}\Sigma_{V}R^{T}_{V}$ be the thin SVDs of $U$ and $V$, respectively, where $L_{U}\!\in\! \mathbb{R}^{m\times d}$, $L_{V}\!\in\! \mathbb{R}^{n\times d}$, and $R_{U},\Sigma_{U},R_{V},\Sigma_{V}\!\in\!\mathbb{R}^{d\times d}$. $X\!=\!L_{X}\Sigma_{X}R^{T}_{X}$, where the columns of $L_{X}\!\in\! \mathbb{R}^{m\times d}$ and $R_{X}\!\in\!\mathbb{R}^{n\times d}$ are the left and right singular vectors associated with the top $d$ singular values of $X$ with rank at most $r$ $(r\!\leq\! d)$, and $\Sigma_{X}\!=\!\textup{diag}([\sigma_{1}(X),\!\cdots\!,\sigma_{r}(X),0,\!\cdots\!,0])\!\in\!\mathbb{R}^{d\times d}$.

Recall that $X\!=\!UV^{T}$, i.e., $L_{X}\Sigma_{X}R^{T}_{X}\!=\!L_{U}\Sigma_{U}R^{T}_{U}R_{V}\Sigma_{V}L^{T}_{V}$, then $\exists\, O_{1},\widehat{O}_{1}\!\in\! \mathbb{R}^{d\times d}$ satisfy $L_{X}\!=\!L_{U}O_{1}$ and $L_{U}\!=\!L_{X}\widehat{O}_{1}$, which implies that $O_{1}\!=\!L^{T}_{U}L_{X}$ and $\widehat{O}_{1}\!=\!L^{T}_{X}L_{U}$. Thus, $O_{1}\!=\!\widehat{O}^{T}_{1}$. Since $L_{X}\!=\!L_{U}O_{1}\!=\!L_{X}\widehat{O}_{1}O_{1}$, we have $\widehat{O}_{1}O_{1}\!=\!O^{T}_{1}O_{1}\!=\!I_{d}$. Similarly, we have $O_{1}\widehat{O}_{1}\!=\!O_{1}O^{T}_{1}\!=\!I_{d}$. In addition, $\exists\, O_{2}\!\in\!\mathbb{R}^{d\times d}$ satisfies $R_{X}\!=\!L_{V}O_{2}$ with $O_{2}O^{T}_{2}\!=\!O^{T}_{2}O_{2}\!=\!I_{d}$. Let $O_{3}\!=\!O_{2}O^{T}_{1}\!\in\! \mathbb{R}^{d\times d}$, then we have $O_{3}O^{T}_{3}\!=\!O^{T}_{3}O_{3}\!=\!I_{d}$, i.e., $\sum_{i}(O_{3})^{2}_{ij}\!=\!\sum_{j}(O_{3})^{2}_{ij}\!=\!1$ for $\forall i,j\in \{1,2,\ldots,d\}$, where $a_{i,j}$ denotes the element of the matrix $A$ in the $i$-th row and the $j$-th column. In addition, let $O_{4}\!=\!R^{T}_{U}R_{V}$, we have $\sum_{i}(O_{4})^{2}_{ij}\!\leq\! 1$ and $\sum_{j}(O_{4})^{2}_{ij}\!\leq\! 1$ for $\forall i,j\in \{1,2,\ldots,d\}$.

According to the above analysis, then we have $O_{2}\Sigma_{X}O^{T}_{2}\!=\!O_{2}O^{T}_{1}\Sigma_{U}O_{4}\Sigma_{V}\!=\!O_{3}\Sigma_{U}O_{4}\Sigma_{V}$. Let $\varrho_{i}$ and $\tau_{j}$ denote the $i$-th and the $j$-th diagonal elements of $\Sigma_{U}$ and $\Sigma_{V}$, respectively. In the following, we consider the two cases of $p_{1}$ and $p_{2}$, i.e., at least one of $p_{1}$ and $p_{2}$ must be no less than 1, or both of them are smaller than 1. It is clear that for any $1/2\!\leq\!p\!\leq\!1$ and $p_{1},p_{2}\!>\!0$ satisfying ${1}/{p_{1}}\!+\!{1}/{p_{2}}\!=\!{1}/{p}$, at least one of $p_{1}$ and $p_{2}$ must be no less than 1. On the other hand, only if $0\!<\!p\!<\!1/2$, there exist $0\!<\!p_{1}\!<\!1$ and $0\!<\!p_{2}\!<\!1$ such that ${1}/{p_{1}}\!+\!{1}/{p_{2}}\!=\!{1}/{p}$, i.e., both of them are smaller than 1.

\textbf{Case 1}. For any $0\!<\!p\!\leq\!1$, there exist $p_{1}\!>\!0$ and $p_{2}\!>\!0$, at least one of which is no less than 1, such that ${1}/{p_{1}}\!+\!{1}/{p_{2}}\!=\!{1}/{p}$.  Without loss of generality, we assume that $p_{2}\!\geq\!1$. Here, we set $k_{1}\!=\!p_{1}/p$ and $k_{2}\!=\!p_{2}/p$. Clearly, we can know that $k_{1},k_{2}>1$ and ${1}/{k_{1}}\!+\!{1}/{k_{2}}\!=\!1$. From Lemma~\ref{lem12}, we have
\begin{equation*}
\begin{split}
\|X\|_{S_{p}}&\leq\left(\textup{Tr}^{p}(O_{2}\Sigma_{X}O^{T}_{2})\right)^{{1}/{p}}=\left(\textup{Tr}^{p}(O_{2}O^{T}_{1}\Sigma_{U}O_{4}\Sigma_{V})\right)^{{1}/{p}}=\left(\textup{Tr}^{p}(O_{3}\Sigma_{U}O_{4}\Sigma_{V})\right)^{{1}/{p}}\\
&=\left(
\sum^{d}_{i=1}\left[\sum^{d}_{j=1}\tau_{j}(O_{3})_{ij}(O_{4})_{ji}\varrho_{i}\right ]^{p}\right )^{{1}/{p}}\\
&=\left(
\sum^{d}_{i=1}\varrho^{p}_{i}\left(\sum^{d}_{j=1}\tau_{j}(O_{3})_{ij}(O_{4})_{ji}\right)^{p}\right )^{{1}/{p}}\\
&^{a}\!\!\leq \left(\left[
\sum^{d}_{i=1} (\varrho^{p}_{i})^{k_{1}}\right]^{{1}/{k_{1}}}\left[
\sum^{d}_{i=1} \left(\sum^{d}_{j=1}\tau_{j}(O_{3})_{ij}(O_{4})_{ji}\right)^{p\times k_{2}}\right]^{{1}/{k_{2}}}\right)^{{1}/{p}}\\
&=\left(\sum^{d}_{i=1}\varrho^{p_{1}}_{i}\right)^{{1}/{p_{1}}}\left[
\sum^{d}_{i=1} \left(\sum^{d}_{j=1}\tau_{j}(O_{3})_{ij}(O_{4})_{ji}\right)^{p_{2}}\right]^{{1}/{p_{2}}}\\
&^{b}\!\!\leq\left(\sum^{d}_{i=1}\varrho^{p_{1}}_{i}\right)^{{1}/{p_{1}}}\left[
\sum^{d}_{i=1} \left(\sum^{d}_{j=1}\tau_{j}\frac{(O_{3})^{2}_{ij}+(O_{4})^{2}_{ji}}{2}\right)^{p_{2}}\right]^{{1}/{p_{2}}}\\
&^{c}\!\!\leq\left(\sum^{d}_{i=1}\varrho^{p_{1}}_{i}\right)^{{1}/{p_{1}}}\left(\sum^{d}_{j=1}\tau^{p_{2}}_{j}\right)^{{1}/{p_{2}}}\\
\end{split}
\end{equation*}
where the inequality $^{a}\!\!\leq$ holds due to the H\"{o}lder's inequality~\cite{mitrinovic:ai}, as stated in Lemma~\ref{lem2}. In addition, the inequality $^{b}\!\!\leq$ follows from the basic inequality $xy\!\leq\! \frac{x^{2}+y^{2}}{2}$ for any real numbers $x$ and $y$, and the inequality $^{c}\!\!\leq$ relies on the facts that $\sum_{i}(O_{3})^{2}_{ij}\!=\!1$ and $\sum_{i}(O_{4})^{2}_{ji}\!\leq\! 1$, and we apply the Jensen's inequality (see Lemma~\ref{lem11}) for the convex function $h(x)=x^{p_{2}}$ with $p_{2}\!\geq\! 1$.

Thus, for any matrices $U\!\in\!\mathbb{R}^{m\times d}$ and $V\!\in\!\mathbb{R}^{n\times d}$ satisfying $X\!=\!UV^{T}$, we have
\begin{equation}\label{pequ1}
\|X\|_{S_{p}}\leq\|U\|_{S_{p_{1}}}\|V\|_{S_{p_{2}}}.
\end{equation}

On the other hand, let $U_{\star}\!=\!L_{X}\Sigma^{{p}/{p_{1}}}_{X}$ and $V_{\star}\!=\!R_{X}\Sigma^{{p}/{p_{2}}}_{X}$, where $\Sigma^{p}_{X}$ is entry-wise power to $p$, then we obtain
\begin{equation*}
X=U_{\star}V^{T}_{\star},\;\;\;\|U_{\star}\|^{p_{1}}_{S_{p_{1}}}=\|V_{\star}\|^{p_{2}}_{S_{p_{2}}}=\|X\|^{p}_{S_{p}}\;\;\;\textrm{with}\;\;\;{1}/{p}={1}/{p_{1}}+{1}/{p_{2}},
\end{equation*}
and
\begin{equation*}
\|X\|_{S_{p}}=\left(\textup{Tr}^{p}(\Sigma_{X})\right)^{1/p}=\|U_{\star}\|_{S_{p_{1}}}\|V_{\star}\|_{S_{p_{2}}}.
\end{equation*}

Therefore, under the constraint $X\!=\!UV^{T}$, we have
\begin{displaymath}
\|X\|_{S_{p}}=\min_{U\in\mathbb{R}^{m\times d},V\in\mathbb{R}^{n\times d}:X=UV^{T}}\|U\|_{S_{p_{1}}}\|V\|_{S_{p_{2}}}.
\end{displaymath}

\textbf{Case 2}. For any $0\!<\! p\!<\! 1/2$, there exist $0\!<\!\widehat{p}_{1}\!<\!1$ and $0\!<\!\widehat{p}_{2}\!<\!1$ such that ${1}/{\widehat{p}_{1}}\!+\!{1}/{\widehat{p}_{2}}\!=\!{1}/{p}$. Next we prove that the result of Theorem~\ref{the11} also holds.
Naturally, for any given $p$, there must exist $p_{1}\!>\!0$ and $p_{2}\!\geq\! 1$ such that ${1}/{p_{1}}\!+\!{1}/{p_{2}}\!=\!{1}/{p}$ and ${1}/{p_{1}}\!=\!{1}/{\widehat{p}_{1}}\!+\!{1}/{q}$ with $q\!\geq\! 1$. Clearly, we can know that ${1}/{\widehat{p}_{1}}\!<\!{1}/{p_{1}}$. Let $U^{*}\!=\!L_{X}\Sigma^{p/p_{1}}_{X}$, $V^{*}\!=\!R_{X}\Sigma^{p/p_{2}}_{X}$, $U^{*}_{1}\!=\!L_{X}\Sigma^{p/\widehat{p}_{1}}_{X}$ and $V^{*}_{1}\!=\!R_{X}\Sigma^{p/\widehat{p}_{2}}_{X}$, then we have
\begin{equation*}
X=U^{*}(V^{*})^{T}=U^{*}_{1}(V^{*}_{1})^{T}
\end{equation*}
from which it follows that
\begin{equation}\label{pequ2}
\|X\|_{S_{p}}=\|U^{*}\|_{S_{p_{1}}}\|V^{*}\|_{S_{p_{2}}}=\|U^{*}_{1}\|_{S_{\widehat{p}_{1}}}\|V^{*}_{1}\|_{S_{\widehat{p}_{1}}}.
\end{equation}

Since ${1}/{p}\!=\!{1}/{p_{1}}\!+\!{1}/{p_{2}}\!=\!{1}/{\widehat{p}_{1}}\!+\!{1}/{\widehat{p}_{2}}$ and ${1}/{p_{1}}\!=\!{1}/{\widehat{p}_{1}}\!+\!{1}/{q}$, then ${1}/{\widehat{p}_{2}}\!=\!{1}/{q}\!+\!{1}/{p_{2}}$. Consider any factor matrices $U$ and $V$ satisfying $X\!=\!UV^{T}$, $V\!=\!L_{V}\Sigma_{V}R^{T}_{V}$ is the thin SVD of $V$. Let $U_{1}\!=\!UU^{T}_{2}$ and $V_{1}\!=\!L_{V}\Sigma^{\widehat{p}_{2}/p_{2}}_{V}$, where $U^{T}_{2}\!=\!R_{V}\Sigma^{\widehat{p}_{2}/q}_{V}$, then it is not difficult to verify that
\begin{equation}\label{pequ4}
\begin{split}
&V=V_{1}U_{2},\, X=U_{1}V^{T}_{1},\\
&\|V\|_{S_{\widehat{p}_{2}}}=\|U_{2}\|_{S_{q}}\|V_{1}\|_{S_{p_{2}}},\\
&\|U_{1}\|_{S_{p_{1}}}\leq\|U\|_{S_{\widehat{p}_{1}}}\|U_{2}\|_{S_{q}}
\end{split}
\end{equation}
where the above inequality follows from \eqref{pequ1} with $q\geq 1$. Combining \eqref{pequ2} and \eqref{pequ4}, for any $U$ and $V$ satisfying $X=UV^{T}$, we have
\begin{equation}\label{pequ41}
\begin{split}
\|X\|_{S_{p}}&=\|U^{*}\|_{S_{p_{1}}}\|V^{*}\|_{S_{p_{2}}}\\
&\leq \|U_{1}\|_{S_{p_{1}}}\|V_{1}\|_{S_{p_{2}}}\\
&\leq \|U\|_{S_{\widehat{p}_{1}}}\|U_{2}\|_{S_{q}}\|V_{1}\|_{S_{p_{2}}}\\
&=\|U\|_{S_{\widehat{p}_{1}}}\|V\|_{S_{\widehat{p}_{2}}}
\end{split}
\end{equation}
where the first inequality follows from \eqref{pequ1}. Recall that
\begin{equation}\label{pequ42}
\|X\|_{S_{p}}=\|U^{*}_{1}\|_{S_{\widehat{p}_{1}}}\|V^{*}_{1}\|_{S_{\widehat{p}_{2}}}.
\end{equation}
Therefore, for any $0\!<\!\widehat{p}_{1}\!<\!1$ and $0\!<\!\widehat{p}_{2}\!<\!1$ satisfying ${1}/{p}\!=\!{1}/{\widehat{p}_{1}}\!+\!{1}/{\widehat{p}_{2}}$, and by \eqref{pequ41} and \eqref{pequ42}, we also have
\begin{equation*}
\|X\|_{S_{p}}=\min_{U\in\mathbb{R}^{m\times d},V\in\mathbb{R}^{n\times d}:X=UV^{T}}\|U\|_{S_{\widehat{p}_{1}}}\|V\|_{S_{\widehat{p}_{2}}}.
\end{equation*}

In summary, for any $0\!<\!p\!\leq\!1$, $p_{1}\!>\!0$ and $p_{2}\!>\!0$ satisfying ${1}/{p}\!=\!{1}/{p_{1}}\!+\!{1}/{p_{2}}$, we have
\begin{equation*}
\|X\|_{S_{p}}=\min_{U\in\mathbb{R}^{m\times d},V\in\mathbb{R}^{n\times d}:X=UV^{T}}\|U\|_{S_{p_{1}}}\|V\|_{S_{p_{2}}}.
\end{equation*}

This completes the proof.
\end{IEEEproof}

\subsection{Proof of Theorem~\ref{cor1}}
\label{proofs2}

\begin{IEEEproof}
Because $p_{1}=p_{2}=2p>0$ and ${1}/{p_{1}}\!+\!{1}/{p_{2}}\!=\!{1}/{p}$, and using Theorem~\ref{the11}, we obtain
\begin{equation*}
\|X\|_{S_{p}}=\min_{U\in\mathbb{R}^{m\times d},V\in\mathbb{R}^{n\times d}:X=UV^{T}}\|U\|_{S_{2p}}\|V\|_{S_{2p}}.
\end{equation*}
Due to the basic inequality $xy\leq \frac{x^2+y^2}{2}$ for any real numbers $x$ and $y$, we have
\begin{equation*}
\begin{split}
\|X\|_{S_{p}}=&\min_{U\in\mathbb{R}^{m\times d},V\in\mathbb{R}^{n\times d}:X=UV^{T}}\|U\|_{S_{2p}}\|V\|_{S_{2p}}\\
=&\min_{U\in\mathbb{R}^{m\times d},V\in\mathbb{R}^{n\times d}:X=UV^{T}}\left(\|U\|^{p}_{S_{2p}}\|V\|^{p}_{S_{2p}}\right)^{1/p}\\
\leq&\min_{U\in\mathbb{R}^{m\times d},V\in\mathbb{R}^{n\times d}:X=UV^{T}}\left(\frac{\|U\|^{2p}_{S_{2p}}\!+\|V\|^{2p}_{S_{2p}}}{2}\right)^{1/p}.
\end{split}
\end{equation*}
Let $U_{\star}\!=\!L_{X}\Sigma^{1/2}_{X}$ and $V_{\star}\!=\!R_{X}\Sigma^{1/2}_{X}$, where $\Sigma^{1/2}_{X}$ is entry-wise power to $1/2$, then we obtain
\begin{equation*}
X=U_{\star}V^{T}_{\star},\;\;\;\|U_{\star}\|^{2p}_{S_{2p}}=\|V_{\star}\|^{2p}_{S_{2p}}=\|X\|^{p}_{S_{p}},
\end{equation*}
which implies that
\begin{equation*}
\|X\|_{S_{p}}=\|U_{\star}\|_{S_{2p}}\|V_{\star}\|_{S_{2p}}=\left(\frac{\|U_{\star}\|^{2p}_{S_{2p}}\!+\|V_{\star}\|^{2p}_{S_{2p}}}{2}\right)^{1/p}.
\end{equation*}
The theorem now follows because
\begin{equation*}
\begin{split}
&\min_{U\in\mathbb{R}^{m\times d},V\in\mathbb{R}^{n\times d}:X=UV^{T}}\|U\|_{S_{2p}}\|V\|_{S_{2p}}\\
=&\min_{U\in\mathbb{R}^{m\times d},V\in\mathbb{R}^{n\times d}:X=UV^{T}}\left(\frac{\|U\|^{2p}_{S_{2p}}\!+\|V\|^{2p}_{S_{2p}}}{2}\right)^{1/p}.
\end{split}
\end{equation*}
This completes the proof.
\end{IEEEproof}

\subsection{Proof of Theorem~\ref{cor3}}
\label{proofs3}

\begin{IEEEproof}
For any $0\!<\!p\!\leq\!1$, $p_{1}\!>\!0$ and $p_{2}\!>\!0$ satisfying ${1}/{p_{1}}\!+\!{1}/{p_{2}}\!=\!{1}/{p}$, and using Theorem~\ref{the11}, we have
\begin{equation*}
\|X\|_{S_{p}}=\min_{U\in\mathbb{R}^{m\times d},V\in\mathbb{R}^{n\times d}:X=UV^{T}}\|U\|_{S_{p_{1}}}\|V\|_{S_{p_{2}}}.
\end{equation*}
Let $k_{1}=(p_{1}\!+\!p_{2})/{p_{2}}$ and $k_{2}=(p_{1}\!+\!p_{2})/{p_{1}}$, we can know that ${1}/{k_{1}}\!+\!{1}/{k_{2}}\!=\!1$. Then
\begin{equation*}
\begin{split}
\|X\|_{S_{p}}&=\min_{U\in\mathbb{R}^{m\times d},V\in\mathbb{R}^{n\times d}:X=UV^{T}}\|U\|_{S_{p_{1}}}\|V\|_{S_{p_{2}}}\\
&=\min_{U\in\mathbb{R}^{m\times d},V\in\mathbb{R}^{n\times d}:X=UV^{T}}\left(\|U\|^{p}_{S_{p_{1}}}\|V\|^{p}_{S_{p_{2}}}\right)^{1/p}\\
&\leq \min_{U\in\mathbb{R}^{m\times d},V\in\mathbb{R}^{n\times d}:X=UV^{T}}\left(\frac{\|U\|^{pk_{1}}_{S_{p_{1}}}}{k_{1}}+\frac{\|V\|^{pk_{2}}_{S_{p_{2}}}}{k_{2}}\right)^{1/p}\\
&=\min_{U\in\mathbb{R}^{m\times d},V\in\mathbb{R}^{n\times d}:X=UV^{T}}\left(\frac{p_{2}\|U\|^{p_{1}}_{S_{p_{1}}}\!\!+p_{1}\|V\|^{p_{2}}_{S_{p_{2}}}}{p_{1}+p_{2}}\right)^{1/p}
\end{split}
\end{equation*}
where the above inequality follows from the well-known Young's inequality, as stated in Lemma~\ref{lem3}, and the monotone increasing property of the function $g(x)=x^{1/p}$.

Let $U_{\star}\!=\!L_{X}\Sigma^{p/p_{1}}_{X}$ and $V_{\star}\!=\!R_{X}\Sigma^{p/p_{2}}_{X}$, then $X=U_{\star}V^{T}_{\star}$. Using Theorem~\ref{the11}, we have
\begin{equation*}
\begin{split}
\|X\|_{S_{p}}&=\min_{U\in\mathbb{R}^{m\times d},V\in\mathbb{R}^{n\times d}:X=UV^{T}}\|U\|_{S_{p_{1}}}\|V\|_{S_{p_{2}}}\\
&=\|U_{\star}\|_{S_{p_{1}}}\|V_{\star}\|_{S_{p_{2}}}\\
&=\left(\frac{p_{2}\|U_{\star}\|^{p_{1}}_{S_{p_{1}}}\!\!+p_{1}\|V_{\star}\|^{p_{2}}_{S_{p_{2}}}}{p_{1}+p_{2}}\right)^{1/p}
\end{split}
\end{equation*}
which implies that
\begin{equation}
\begin{split}
\|X\|_{S_{p}}&=\min_{U\in\mathbb{R}^{m\times d},V\in\mathbb{R}^{n\times d}:X=UV^{T}}\|U\|_{S_{p_{1}}}\|V\|_{S_{p_{2}}}\\
&=\min_{U\in\mathbb{R}^{m\times d},V\in\mathbb{R}^{n\times d}:X=UV^{T}}\left(\frac{p_{2}\|U\|^{p_{1}}_{S_{p_{1}}}\!\!+p_{1}\|V\|^{p_{2}}_{S_{p_{2}}}}{p_{1}+p_{2}}\right)^{1/p}\\
&=\min_{U\in\mathbb{R}^{m\times d},V\in\mathbb{R}^{n\times d}:X=UV^{T}}\left(\frac{\|U\|^{p_{1}}_{S_{p_{1}}}/p_{1}\!+\|V\|^{p_{2}}_{S_{p_{2}}}/p_{2}}{1/p}\right)^{1/p}.
\end{split}
\end{equation}

This completes the proof.
\end{IEEEproof}

\subsection{Proof of Theorem~\ref{the12}}
\label{proofs4}

\begin{IEEEproof}
Let $U\!\in\!\mathbb{R}^{m\times d}$ and $\widehat{V}\!\in\!\mathbb{R}^{n\times d}$ be any factor matrices such that $X\!=\!U\widehat{V}^{T}$, and $\widehat{p}_{1}\!=\!p_{1}\!>\!0$ and $\widehat{p}_{2}\!=\!p_{2}p_{3}/(p_{2}\!+\!p_{3})\!>\!0$, which means that $1/\widehat{p}_{1}\!+\!1/\widehat{p}_{2}\!=\!1/p$. According to Theorem~\ref{the11}, we obtain
\begin{equation}\label{equ20}
\|X\|_{S_{p}}=\min_{U\in\mathbb{R}^{m\times d},\widehat{V}\in\mathbb{R}^{n\times d}:X=U\widehat{V}^{T}}\|U\|_{S_{\widehat{p}_{1}}}\|\widehat{V}\|_{S_{\widehat{p}_{2}}}.
\end{equation}

Let $V\!\in\!\mathbb{R}^{d\times d}$ and $W\!\in\!\mathbb{R}^{n\times d}$ be factor matrices of $\widehat{V}$, i.e., $VW^{T}\!=\!\widehat{V}^{T}$. Since $\widehat{p}_{2}\!=\!p_{2}p_{3}/(p_{2}\!+\!p_{3})$, then $1/\widehat{p}_{2}\!=\!1/p_{2}\!+\!1/p_{3}$. Using Theorem~\ref{the11}, we also have
\begin{equation}\label{equ21}
\|\widehat{V}\|_{S_{\widehat{p}_{2}}}=\min_{V\in\mathbb{R}^{d\times d},W\in\mathbb{R}^{n\times d}:\widehat{V}=(VW^{T})^{T}}\|V\|_{S_{p_{2}}}\|W\|_{S_{p_{3}}}.
\end{equation}
Combining \eqref{equ20} and \eqref{equ21}, we obtain
\begin{equation*}
\|X\|_{S_{p}}=\min_{U\in\mathbb{R}^{m\times d},V\in\mathbb{R}^{d\times d},W\in\mathbb{R}^{n\times d}:X=UVW^{T}}\|U\|_{S_{p_{1}}}\|V\|_{S_{p_{2}}}\|W\|_{S_{p_{3}}}.
\end{equation*}

Using the above result, we have
\begin{equation*}
\begin{split}
\|X\|_{S_{p}}=&\min_{U\in\mathbb{R}^{m\times d},V\in\mathbb{R}^{d\times d},W\in\mathbb{R}^{n\times d}:X=UVW^{T}}\|U\|_{S_{p_{1}}}\|V\|_{S_{p_{2}}}\|W\|_{S_{p_{3}}}\\
=&\min_{U\in\mathbb{R}^{m\times d},V\in\mathbb{R}^{d\times d},W\in\mathbb{R}^{n\times d}:X=UVW^{T}}\left(\|U\|^{p}_{S_{p_{1}}}\|V\|^{p}_{S_{p_{2}}}\|W\|^{p}_{S_{p_{3}}}\right)^{1/p}\\
\leq&\min_{U\in\mathbb{R}^{m\times d},V\in\mathbb{R}^{d\times d},W\in\mathbb{R}^{n\times d}:X=UVW^{T}} \left(\frac{p_{2}p_{3}\|U\|^{p_{1}}_{S_{p_{1}}}\!\!+p_{1}p_{3}\|V\|^{p_{2}}_{S_{p_{2}}}\!\!+p_{1}p_{2}\|W\|^{p_{3}}_{S_{p_{3}}}}{p_{2}p_{3}+p_{1}p_{3}+p_{1}p_{2}}\right)^{1/p}
\end{split}
\end{equation*}
where the above inequality follows from the well-known Young's inequality, as stated in Lemma~\ref{lem3}.

Let $U_{\star}\!=\!L_{X}\Sigma^{p/p_{1}}_{X}$, $V_{\star}\!=\!\Sigma^{p/p_{2}}_{X}$ and $W_{\star}\!=\!R_{X}\Sigma^{p/p_{3}}_{X}$, it is easy to verify that $X=U_{\star}V_{\star}W^{T}_{\star}$. Then we have
\begin{equation*}
\begin{split}
\|X\|_{S_{p}}=&\min_{U\in\mathbb{R}^{m\times d},V\in\mathbb{R}^{d\times d},W\in\mathbb{R}^{n\times d}:X=UVW^{T}}\|U\|_{S_{p_{1}}}\|V\|_{S_{p_{2}}}\|W\|_{S_{p_{3}}}\\
=&\:\|U_{\star}\|_{S_{p_{1}}}\|V_{\star}\|_{S_{p_{2}}}\|W_{\star}\|_{S_{p_{3}}}\\
=&\left(\frac{p_{2}p_{3}\|U_{\star}\|^{p_{1}}_{S_{p_{1}}}\!\!+p_{1}p_{3}\|V_{\star}\|^{p_{2}}_{S_{p_{2}}}\!\!+p_{1}p_{2}\|W_{\star}\|^{p_{3}}_{S_{p_{3}}}}{p_{2}p_{3}+p_{1}p_{3}+p_{1}p_{2}}\right)^{1/p}.
\end{split}
\end{equation*}

Therefore, we have
\begin{equation*}
\begin{split}
\|X\|_{S_{p}}=&\min_{U\in\mathbb{R}^{m\times d},V\in\mathbb{R}^{d\times d},W\in\mathbb{R}^{n\times d}:X=UVW^{T}}\|U\|_{S_{p_{1}}}\|V\|_{S_{p_{2}}}\|W\|_{S_{p_{3}}}\\
=&\min_{U\in\mathbb{R}^{m\times d},V\in\mathbb{R}^{d\times d},W\in\mathbb{R}^{n\times d}:X=UVW^{T}} \left(\frac{p_{2}p_{3}\|U\|^{p_{1}}_{S_{p_{1}}}\!\!+p_{1}p_{3}\|V\|^{p_{2}}_{S_{p_{2}}}\!\!+p_{1}p_{2}\|W\|^{p_{3}}_{S_{p_{3}}}}{p_{2}p_{3}+p_{1}p_{3}+p_{1}p_{2}}\right)^{1/p}\\
=&\min_{U\in\mathbb{R}^{m\times d},V\in\mathbb{R}^{d\times d},W\in\mathbb{R}^{n\times d}:X=UVW^{T}} \left(\frac{\|U\|^{p_{1}}_{S_{p_{1}}}\!/p_{1}+\|V\|^{p_{2}}_{S_{p_{2}}}\!/p_{2}+\|W\|^{p_{3}}_{S_{p_{3}}}\!/p_{3}}{1/p}\right)^{1/p}.
\end{split}
\end{equation*}

This completes the proof.
\end{IEEEproof}

\subsection{Proof of Corollary~\ref{cor4}}
\label{proofs5}

\begin{IEEEproof}
Since $p_{1}\!=\!p_{2}\!=\!p_{3}\!=\!3p\!>\!0$ and ${1}/{p_{1}}\!+\!{1}/{p_{2}}\!+\!{1}/{p_{3}}\!=\!{1}/{p}$, and using Theorem~\ref{the12}, we have
\begin{equation*}
\|X\|_{S_{p}}=\min_{U\in\mathbb{R}^{m\times d},V\in\mathbb{R}^{d\times d},W\in\mathbb{R}^{n\times d}:X=UVW^{T}}\|U\|_{S_{3p}}\|V\|_{S_{3p}}\|W\|_{S_{3p}}.
\end{equation*}

From the basic inequality $xyz\leq \frac{x^3+y^3+z^3}{3}$ for arbitrary positive numbers $x$, $y$ and $z$, we obtain
\begin{equation*}
\begin{split}
\|X\|_{S_{p}}=&\min_{U\in\mathbb{R}^{m\times d},V\in\mathbb{R}^{d\times d},W\in\mathbb{R}^{n\times d}:X=UVW^{T}}\|U\|_{S_{3p}}\|V\|_{S_{3p}}\|W\|_{S_{3p}}\\
=&\min_{U\in\mathbb{R}^{m\times d},V\in\mathbb{R}^{d\times d},W\in\mathbb{R}^{n\times d}:X=UVW^{T}}\left(\|U\|^{p}_{S_{3p}}\|V\|^{p}_{S_{3p}}\|W\|^{p}_{S_{3p}}\right)^{1/p}\\
\leq&\min_{U\in\mathbb{R}^{m\times d},V\in\mathbb{R}^{d\times d},W\in\mathbb{R}^{n\times d}:X=UVW^{T}}\left(\frac{\|U\|^{3p}_{S_{3p}}\!+\|V\|^{3p}_{S_{3p}}\!+\|W\|^{3p}_{S_{3p}}}{3}\right)^{1/p}.
\end{split}
\end{equation*}
Let $U_{\star}\!=\!L_{X}\Sigma^{1/3}_{X}$, $V_{\star}\!=\!\Sigma^{1/3}_{X}$ and $W_{\star}\!=\!R_{X}\Sigma^{1/3}_{X}$, where $\Sigma^{1/3}_{X}$ is entry-wise power to $1/3$, then we have
\begin{equation*}
X=U_{\star}V_{\star}W^{T}_{\star},\;\;\|U_{\star}\|^{3p}_{S_{3p}}=\|V_{\star}\|^{3p}_{S_{3p}}=\|W_{\star}\|^{3p}_{S_{3p}}=\|X\|^{p}_{S_{p}},
\end{equation*}
which implies that
\begin{equation*}
\|X\|_{S_{p}}=\|U_{\star}\|_{S_{3p}}\|V_{\star}\|_{S_{3p}}\|W_{\star}\|_{S_{3p}}=\left(\frac{\|U_{\star}\|^{3p}_{S_{3p}}\!+\|V_{\star}\|^{3p}_{S_{3p}}\!+\|W_{\star}\|^{3p}_{S_{3p}}}{3}\right)^{1/p}.
\end{equation*}
The theorem now follows because
\begin{equation*}
\begin{split}
&\min_{U\in\mathbb{R}^{m\times d},V\in\mathbb{R}^{d\times d},W\in\mathbb{R}^{n\times d}:X=UVW^{T}}\|U\|_{S_{3p}}\|V\|_{S_{3p}}\|W\|_{S_{3p}}\\
=&\min_{U\in\mathbb{R}^{m\times d},V\in\mathbb{R}^{d\times d},W\in\mathbb{R}^{n\times d}:X=UVW^{T}}\left(\frac{\|U\|^{3p}_{S_{3p}}\!+\|V\|^{3p}_{S_{3p}}\!+\|W\|^{3p}_{S_{3p}}}{3}\right)^{1/p}.
\end{split}
\end{equation*}
This completes the proof.
\end{IEEEproof}

\section{Conclusions}
In general, the SQNM is non-convex, non-smooth and even non-Lipschitz. Most existing algorithms are too slow or even impractical for large-scale problems, due to the SVD or EVD of the whole matrix in each iteration. Therefore, it is very important that how to transform such challenging problems into a simpler one, such as a smooth optimization problem. In this paper, we first presented and rigorously proved that for any $p$, $p_{1}$, $p_{2}\!>\!0$ satisfying $1/p\!=\!1/p_{1}\!+\!1/p_{2}$,  the Schatten-$p$ quasi-norm of any matrix is equivalent to the minimization of the product (or the weighted sum) of the Schatten-$p_{1}$ norm (or quasi-norm) and Schatten-$p_{2}$ norm (or quasi-norm) of two factor matrices. Especially, when $p\!>\!1/2$, there is an equivalence between the Schatten-$p$ quasi-norm (or norm) of any matrix and the Schatten-$2p$ norms of its two factor matrices, e.g., $\|X\|_{*}\!=\!\min_{X=UV^{T}}(\|U\|^{2}_{F}\!+\!\|V\|^{2}_{F})/{2}$. That is, various SQNM problems with $p\!>\!1/2$ can be transformed into a simpler one only involving the smooth norms of two factor matrices, which can naturally lead to simpler and more efficient algorithms than conventional methods.

We further extended the equivalence relationship of two factor matrices to the cases of three and more factor matrices, from which we can see that for any $0\!<\!p\!<\!1$, the Schatten-$p$ quasi-norm of any matrix is the minimization of the mean of the Schatten-$(\lfloor1/p\rfloor\!+\!1)p$ norms of all factor matrices. In other words, for any $0\!<\!p\!<\!1$, the SQNM can be transformed into an optimization problem only involving the smooth norms of multiple factor matrices. Finally, we provided some representative examples for two and three factor matrices. It is clear that the bi-nuclear and Frobenius/nuclear quasi-norms defined in our previous paper~\cite{shang:snm} and the tri-nuclear quasi-norm defined in our previous paper~\cite{shang:tsnm} are three important special cases.

\bibliographystyle{IEEEtran}
\bibliography{IEEEabrv,shang}

\end{document}